\documentclass{SPW9_proc}

\usepackage{shortvrb}
\MakeShortVerb{\|}
\def\pdfLaTeX{pdf\kern.06em\LaTeX}
\begin{document}

\title{The "Second Solar Spectrum" of the post-AGB binary 89 Herculis.}

\author[1,*]{M. Gangi}
\author[2]{F. Leone}

\affil[1]{INAF - Osservatorio Astrofisico di Catania, Via S. Sofia 78, I-95123 Catania, Italy}
\affil[2]{Universit\'a di Catania, Dipartimento di Fisica e Astronomia, Sezione Astrofisica Via S. Sofia 78, I-95123 Catania, Italy}
\affil[*]{\textit{currently at:} INAF - Osservatorio Astronomico di Roma, Via Frascati 33, I-00078 Monte Porzio Catone, Italy.
\textit{Email:} manuele.gangi@inaf.it}

\runningtitle{The "Second Solar Spectrum" of the post-AGB binary 89 Herculis.}
\runningauthor{M. Gangi et al.}

\firstpage{1}

\maketitle

\begin{abstract}
We report a spectropolarimetric study of the post-AGB binary system 89 Herculis, based on data acquired with the high-resolution Catania Astrophysical Observatory Spectropolarimeter, the HArps-North POlarimeter and the Echelle SpectroPolarimetric Device for the Observation of Stars.

The linear polarization clearly detected across single atmospheric lines in absorption is characterized by complex Q and U morphologies that are variable with the orbital period of the system. Gauss-level magnetic fields, continuum depolarization due to pulsations, hot spots and scattering in the circumstellar environment were excluded as possible origin of the observed polarization.

In the context of the optical pumping mechanism, we suggest that the anisotropy of the stellar radiation field, fuelled  by the close binary companion, can be responsible of the observed periodic properties of the spectral line polarization.

We conclude that high resolution linear spectropolarimetry could be an important diagnostic tool in the study of aspherical envelopes of cool and evolved stars.

\end{abstract}

\section{Introduction}
In stellar astrophysics, high resolution linear spectropolarimetry is considered as a promising searching tool for a variety of environments. Despite the difficulties involved in data acquisition and interpretation, it deserves to be a standard observational technique. Indeed, different spectropolarimetric studies has been carry out on evolved stars like the Mira $\rm \chi$ Cygni \citep{Lebre2014,Lopez2019} and the red supergiant Betelgeuse \citep{Auriere2016,Lopez2018}. In analogy to the linearly polarized spectrum of the Sun \citep[the so-called \emph{Second Solar Spectrum},][]{Ivanov1991} it appears that evolved stars can be characterized by the presence of polarization signals across spectral lines. In this context we focused on the polarimetric properties of the post-AGB 89 Herculis.

89 Herculis is considered a prototype of post-AGB binaries with circumbinary disk. This F-type star has an orbital period of about 288 days \citep{Waters1993} and pulsates in 63 days \citep{Ferro1984}. Due to the strong mass loss from the primary star, there are two nebular components: an hour-glass structure and an unresolved circumbinary Keplerian disk \citep{Bujarrabal2007}.

\section{Observational data}
We undertaken a spectropolarimetric observational campaign using the high-resolution Catania Astrophysical Observatory Spectropolarimeter \citep[CAOS: R=55000,][]{Leone2016a} and the HArps-North POlarimeter \citep[HANPO: R=115000,][]{Leone2016b}. Data were reduced and Stokes $\rm Q/I$ and $\rm U/I$ spectra extracted according to \citet{Leone2016a}. Further data, acquired with the Echelle SpectroPolarimetric Device for the Observation of Stars \citep[ESPaDOnS: R=68000,][]{Donati2006} were retrieved from the Canadian Astronomy Data Centre. A total of 26 spectra, covering the full orbital period of the sytem, has been obtained.

\begin{figure}[t]
\centering
\includegraphics[trim= 17cm 12cm 2.5cm 1cm, clip=true,width=6cm,angle=180]{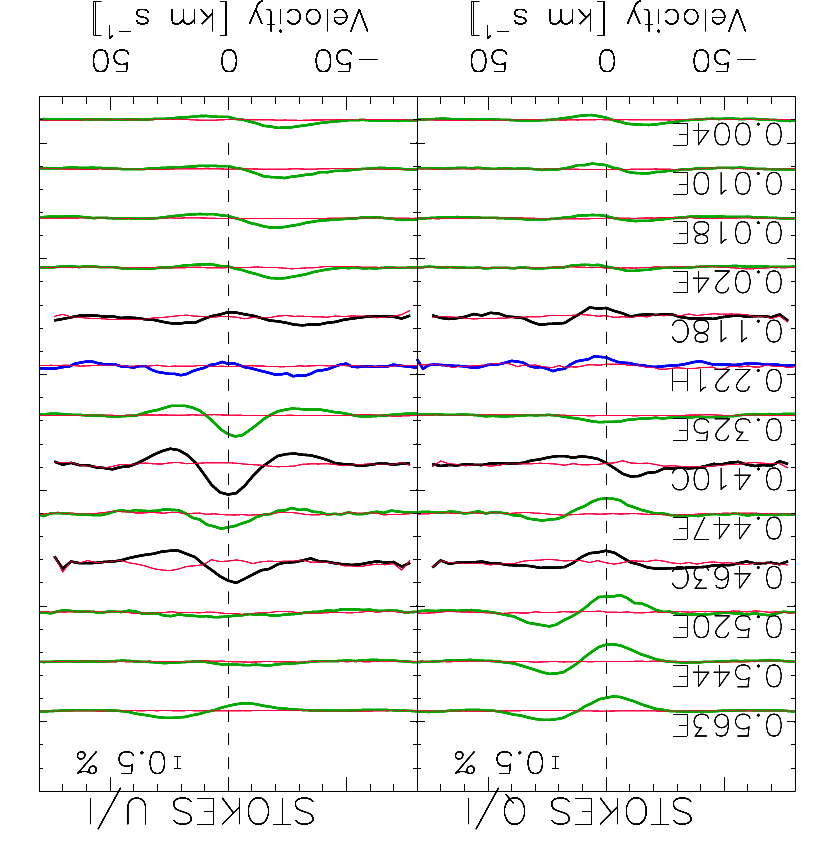}
\includegraphics[trim= 17cm 12cm 2.5cm 1cm, clip=true,width=6cm,angle=180]{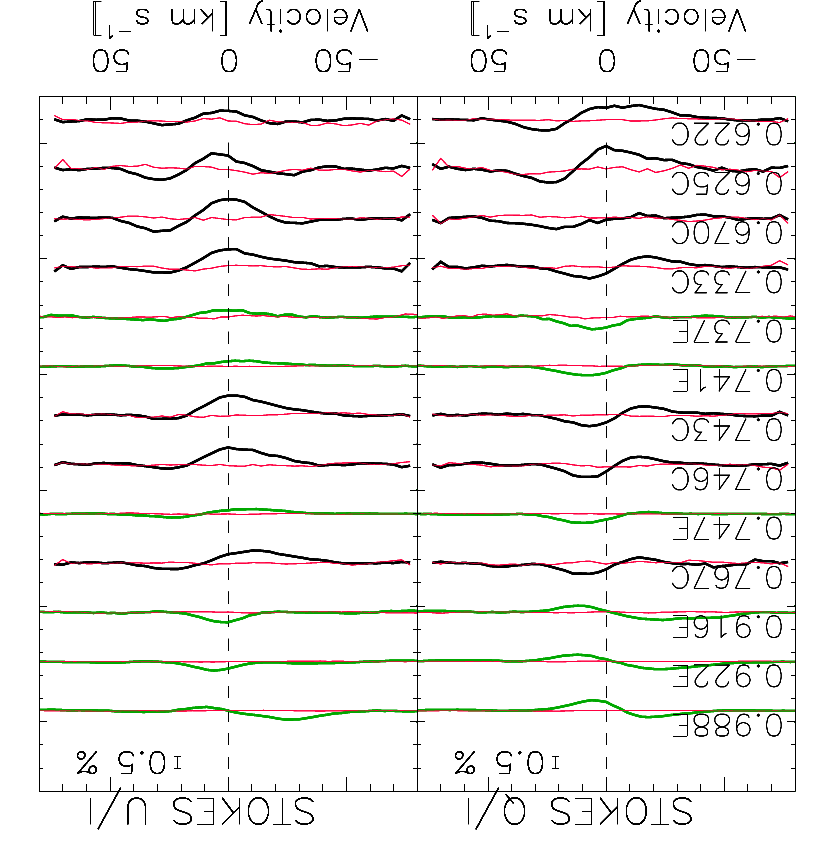}
\begin{center}\caption{\label{fig:LSD_profiles} LSD Stokes $\rm Q/I$ and $\rm U/I$ profiles of 89 Her, vertically shifted for better visualization according to the orbital phase. The corresponding null spectra are shown in red. The null spectra \citep{Leone2016a} check for the presence of any spurious contribution to the polarized spectra and errors in the data reduction process.  'E': ESPaDOnS; 'C': CAOS; 'H': HANPO.}\end{center}
\end{figure}

Despite the photon noise, we have found that the ESPaDOnS $\rm Q/I$ and $\rm U/I$ spectra present not-null signals, that are clearly visible across individual photospheric lines. An Atlas of such spectra is reported in \citet{Gangi2019}. 
Due to the lower resolution, in CAOS spectra we did not found any direct evidence of polarization. For this reason we applied the multi-line LSD technique  \citep{Donati1997} to computed, for each observational date, a very high ($\rm 10^3-10^4$) signal-to-noise ratio (SNr) $\rm I$, $\rm Q/I$ and $\rm U/I$ profile. We have first constructed the line mask from a synthetic spectrum computed by \textsc{synthe} \citep{Kurucz2005} with stellar parameters given in \citet{Waters1993}. Spectral region was chosen between $4250$ \AA\ and $8000$ \AA. Lines weaker than the noise level were excluded, as well as specific elements like $\rm H$, $\rm Ca$ $\rm I$, $\rm Na$ $\rm I$ and telluric lines. A set of about $1000$ spectral lines was then selected for the calculation. Further details about the LSD computation are in \citet{Leone2018}. The obtained LSD profiles show complex $\rm Q/I$ and $\rm U/I$ morphologies variable in time (Fig. \ref{fig:LSD_profiles}).

\section{Origin of polarization}
\begin{figure}[]
\centering
\includegraphics[trim= 16cm 13cm 2.6cm 2.2cm, clip=true,width=4.9cm,angle=180]{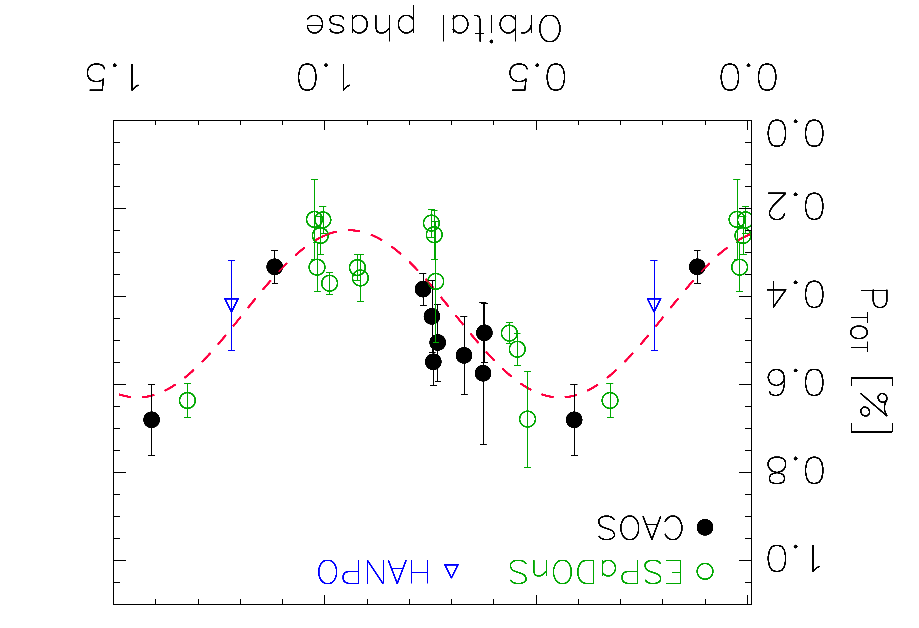}
\includegraphics[trim= 16cm 13cm 2.6cm 2.2cm, clip=true,width=4.9cm,angle=180]{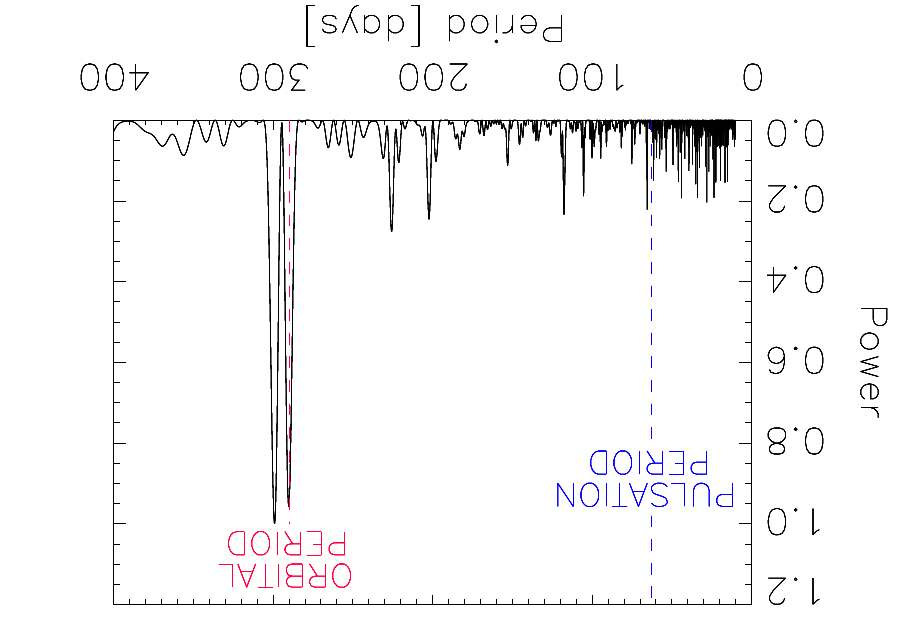}
\begin{center}\caption{\label{fig:P_variability} Left: Total LSD polarization $\rm P=\int{\sqrt{(Q/I)^2+(U/I)^2}d\lambda}$ folded with the orbital phase. The latter was computed according to the following ephemeris: $\rm JD = 2446013.72(\pm16.95)+288.36(\pm0.71)$ days \citep{Waters1993}; a sine fit of variability is also shown. Right: the Lomb-Scargle \citep{Scargle1982} periodogram of P rules out a possible variability with the pulsation period of 89 Her.}\end{center}
\end{figure}

\begin{figure}[]
\centering
\includegraphics[trim= 16cm 13cm 2.6cm 2.2cm, clip=true,width=4.9cm,angle=180]{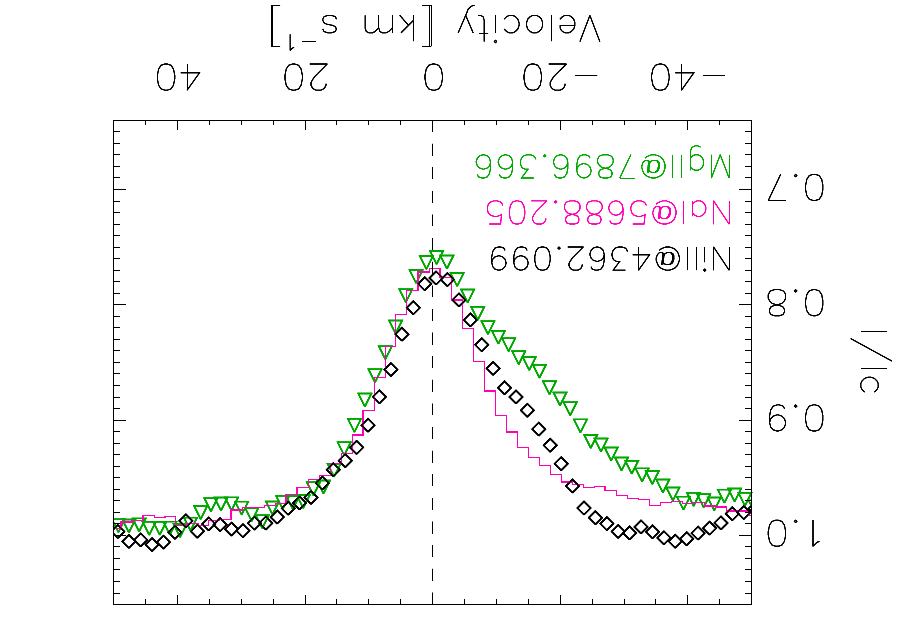}
\includegraphics[trim= 16cm 13cm 2.6cm 2.2cm, clip=true,width=4.9cm,angle=180]{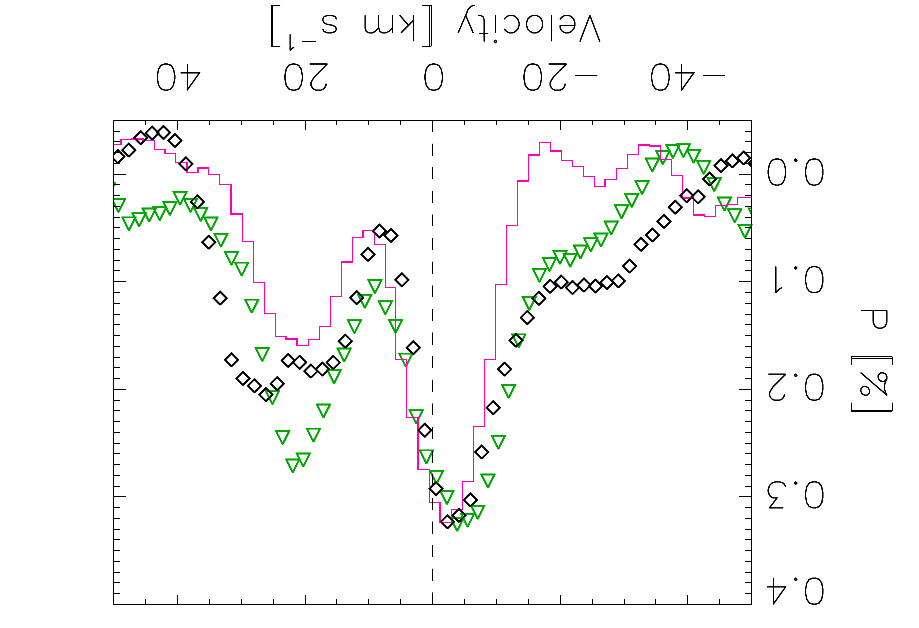}
\begin{center}\caption{\label{fig:same_polarization} Examples of spectral lines with equal depth in different spectral regions (left) showing that, within the errors, polarization does not depend on wavelength (right).}\end{center}
\end{figure}

Spectral lines in absorption with linearly polarized profiles are not common and different phenomena can potentially contribute to the polarization. To find out the mechanisms compatible with our data, we summarize here the main properties of the observed polarization:

\begin{itemize}
\item Temporal variability: the polarization degree measured across the LSD profile of 89 Herculis is variable with the orbital period, while no statistically relevant variability with the pulsation period is found (Fig. \ref{fig:P_variability});
\item Wavelength dependence: the polarization degree of 89 Her is independent of wavelength (Fig. \ref{fig:same_polarization});
\item Circular polarization: 89 Her is a circularly unpolarized source \citep{Sabin2015}. We confirmed this result with new CAOS measurements.
\end{itemize}

In addition, using broad-band optical photopolarimetry, \citet{Akras2017} has found that the continuum flux of 89 Herculis is unpolarized.

From this, we exclude Gauss-level magnetic fields, continuum polarization due to pulsations, hot spots and scattering in the circumstellar environment as the possible origin of the observed polarization.

\section{A possible mechanism: the optical pumping model}

Most of the properties of the linearly polarized spectrum of the Sun \citep{Stenflo1997} are considerably influenced by the so-called optical pumping mechanism \citep{Landi1998}. This mechanism is based on the presence of an anisotropic radiation pumping process, which induce population imbalance between atomic sublevels \citep[e.g.,][]{Truglio1997}. 

In 89 Her, the very close \citep[0.31 AU,][]{Bujarrabal2007} companion star can be responsible for an anisotropy in the total radiation field of the system. Atoms can be then radiatively aligned along the direction of anisotropy, i.e. the primary-secondary direction, which in turn can explain the observed variability with the orbital period.

To numerically explore this possibility we have first computed the anisotropy factor $\rm \omega$ \citep{Landi2004}, which is zero for an isotropic radiation field and 1 for an unidirectional radiation beam. Considering the photospheric and orbital parameters given in \citet{Waters1993} and \citet{Bujarrabal2007}, we have found $\rm \omega \sim 0.8$, that is compatible with the presence of a non-isotropic field between the two components of 89 Her.

\begin{figure}[!h]
\centering
\includegraphics[trim= 16cm 13cm 2.6cm 2.2cm, clip=true,width=4cm,angle=180]{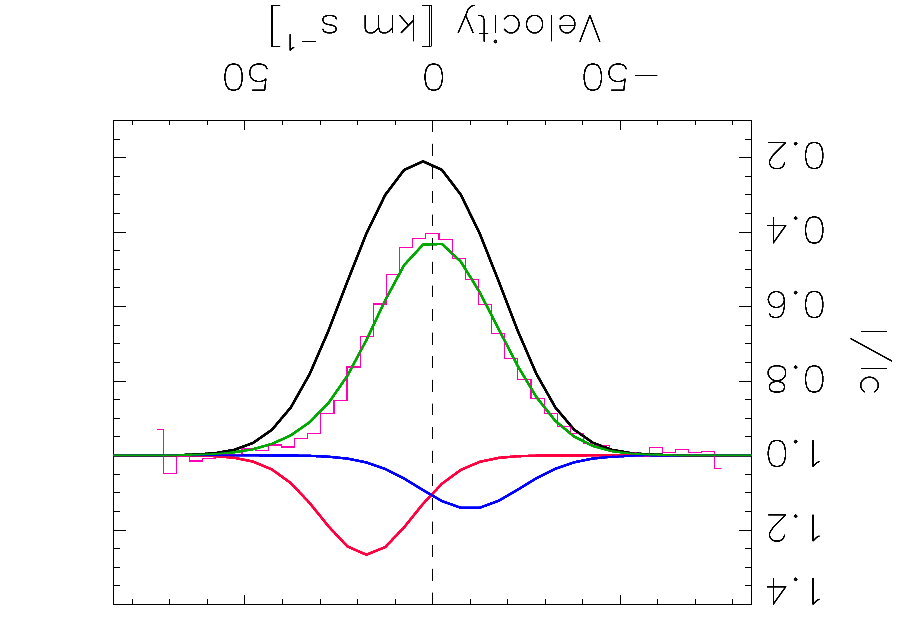}
\includegraphics[trim= 16cm 13cm 2.6cm 2.2cm, clip=true,width=4cm,angle=180]{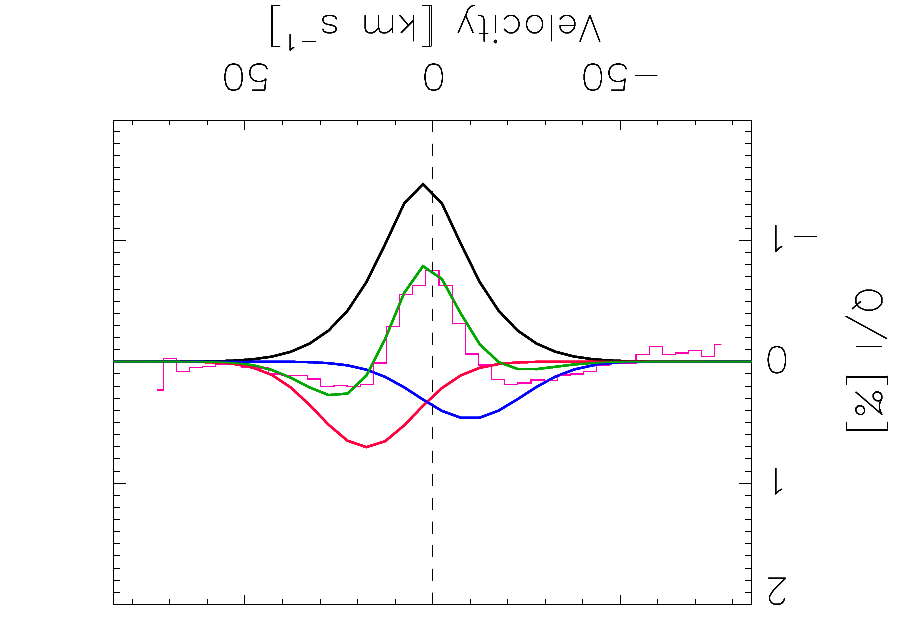}
\includegraphics[trim= 16cm 13cm 2.6cm 2.2cm, clip=true,width=4cm,angle=180]{56816_CAOS_lsd_Q.pdf}
\begin{center}\caption{\label{fig:ex_fit} Example of CAOS-LSD profile (pink) fitted with the \textsc{hazel} code. The solid (grren) continuum is the sum the three components (blue, red and black continuum).}\end{center}
\end{figure}

We have then used the \textsc{hazel} code \citep{Asensio2008} to reproduce the observed Stokes profiles; computation details are in \citet{Leone2018}. We have found that, for each profile, three components are required to fit the Stokes morphologies: a stationary feature in absorption (hereafter \textsc{component 2}) and two oppositely Doppler-shifted components, the blue-shifted \textsc{component 1} and the red-shifted \textsc{component 3} (Fig. \ref{fig:ex_fit}).

\begin{figure}[!h]
\centering
\includegraphics[trim= 17.8cm 13cm 3.8cm 2.2cm, clip=true,width=3.2cm,angle=180]{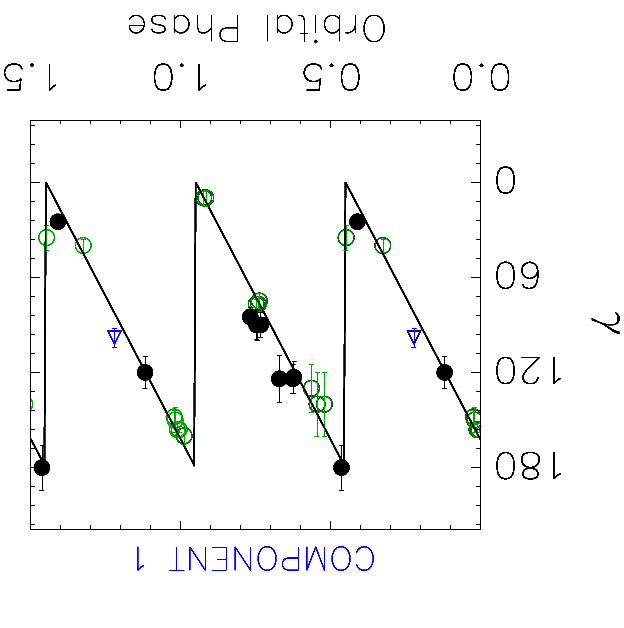}
\includegraphics[trim= 17.8cm 13cm 3.8cm 2.2cm, clip=true,width=3.2cm,angle=180]{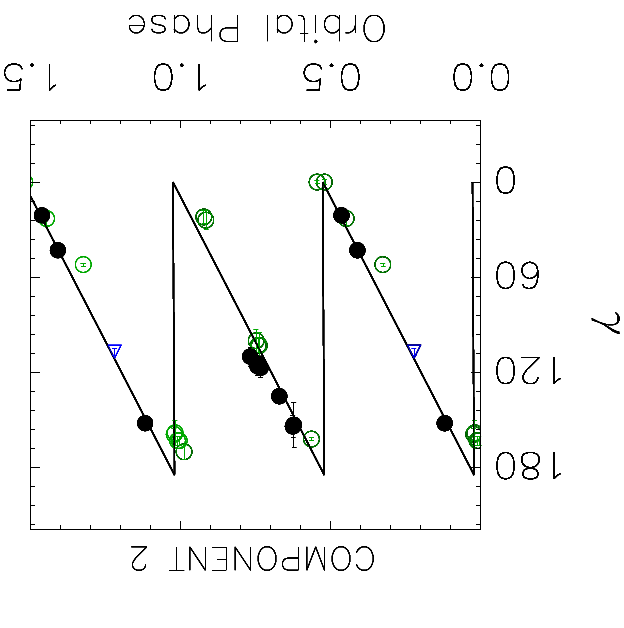}
\includegraphics[trim= 17.8cm 13cm 3.8cm 2.2cm, clip=true,width=3.2cm,angle=180]{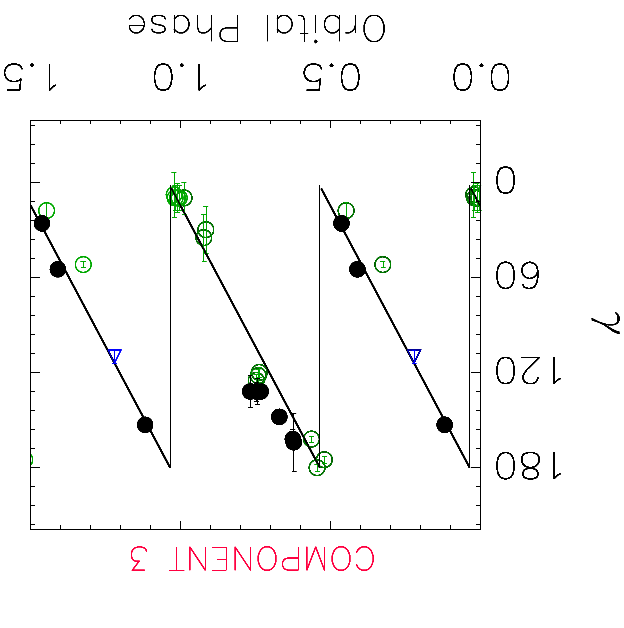}
\begin{center}\caption{\label{fig:results_fit} \textsc{hazel} polarization angles of the line components best matching the Stokes LSD profiles phase-folded with the orbital period. Symbols are as in Fig. \ref{fig:P_variability}.}\end{center}
\end{figure}

Fig. \ref{fig:results_fit} shows the behaviour of the polarization angle $\rm \gamma$ of the three components derived from the \textsc{hazel} fit. The well defined saw-tooth variation of the $\rm \gamma$ angle, due to its definition in the $\rm [0,180^\circ]$ range, is indicative of a closed loop of the polarization vector along the orbital motion of the secondary star. Furthermore, the longitude of periastron $\rm \omega=395.3^\circ$ \citep{Waters1993} implies that the polarization vector is north-south oriented ($\rm \gamma=0^\circ$) in conjunctions (i.e. at phases $\rm \phi=0.5,1.0$) and east-west oriented ($\rm \gamma=90^\circ$) in quadrature (i.e. at phases $\rm \phi=0.25,0.75$).

\section{Discussion and conclusions}
We have found that the optical pumping mechanisms can be at the origin of the observed linear polarization in the post-AGB binary 89 Her. The radiation field anisotropy, which is necessary for the operation of such process, may have origin in the close orbiting star. 

We can qualitatively explain the behaviour of the Stokes morphologies considering the \textsc{component 2} representative of the average optical and physical properties of the primary component in reflecting and reprocessing radiation from the secondary star, while the blue-shifted \textsc{component 1} and red-shifted \textsc{component 3} represent the jet pointing towards us and the receding jet respectively.

This result shows how evolved stars can be characterized by a \emph{Second Solar Spectrum}-like behaviour and gives us new instrument for better understanding the physical mechanisms underlying these kind of systems.

\begin{acknowledgements}

This study is based on observations made with the Catania Astrophysical Observatory Spectropolarimeter (CAOS) operated by the Catania Astrophysical Observatory.

I wish to express my gratitude to Dr. C. Scalia for his support during observations.

\end{acknowledgements}


\end{document}